\documentclass[twocolumn,showpacs,preprintnumbers,amsmath,amssymb,prb]{revtex4}

\usepackage{graphicx}
\usepackage{bm}

\begin{document}
\title{Radial-breathing-like phonon modes of double-walled carbon nanotubes}

\author{Gang Wu}
\email{E-mail: wugaxp@gmail.com}
\author{Jian Zhou}
\author{Jinming Dong}
\email[Corresponding author E-mail: ]{jdong@nju.edu.cn}
\affiliation{Group of Computational Condensed Matter Physics,
National Laboratory of Solid State Microstructures and Department
of Physics, Nanjing University, Nanjing 210093, P. R. China}

\begin{abstract}
The radial-breathing-like phonon modes (RBLMs) of the
double-walled carbon nanotubes are studied in a simple analytical
model, in which the interaction force constants (FCs) can be
obtained analytically from the continuous model. The RBLMs
frequencies are obtained by solving the dynamical matrix, and
their relationship with the tube radii can be obtained
analytically, offering a powerful experimental tool for
determining precisely the radii of the multi-walled carbon
nanotubes.

\end{abstract}

\pacs { 63.22.+m, 78.30.Na}

\date{\today}
\maketitle

Right now, the double-walled carbon nanotubes (DWCNTs) become an
important research subject after they are successfully synthesized
by catalytic chemical vapor deposition \cite{r1,r2,r3} and the
thermal conversion of C$_{60}$ encapsulated in single-walled
carbon nanotubes (SWCNTs) \cite{r4}. Many researches
\cite{r2,r3,r4,r5,r6,r7,r8,r9,r10,r11,r12,r13} have shown that
there exists a kind of radial motion in the DWCNTs similar to the
radial breathing modes (RBMs) in the SWCNTs, called as the radial
breathing-like modes (RBLMs). But their relationship with the tube
radii is still an open question, which is the most important for
understanding the experimental RBLMs. In some publications
\cite{r5,r6,r7,r8,r9,r10,r12}, the force constants are computed
numerically, making the relationship between the RBLM frequencies
and the tube radii unclear. Dobard\v{z}i\'c \textit{et al.}
suggested a simple analytical model\cite{r13}, considering the
tube walls as coupled oscillators, and the FC values of the
coupled oscillators are determined by a fitting to the
experimental B$_{2g}$ graphite phonon frequency of 127 cm$^{ -
1}$. With this model, they have successfully calculated the RBLM
frequencies of the DWCNT.

In this work, we have investigated further the RBLMs of the
DWCNTs, using the simple analytical model \cite{r13}, and assuming
the carbon atoms on the nanotube wall are distributed
continuously, which has been used in a lot of systems
\cite{r14,r20,r23,r24,r25,r26}. Based on the continuum model in
Ref. \onlinecite{r26}, we have derived a set of formulae to
describe the interaction between an infinite length tube and an
atom in any place. With these formulae, the force constants
between the inner and outer tubes can be obtained analytically,
and furthermore, an accurate relationship between the RBLM
frequencies and the tube radii has been obtained. Our final
results only involve a few basic parameter values relying on the
nanotubes structure and the parameters of Lennard-Jones (LJ)
potential.

Now, let us introduce the simple analytical model. For the RBLMs,
the only degree of freedom is the motion along the radial
direction of the tube, so we can simply consider the RBLMs as the
motion of a harmonic oscillator. A segment of SWNT can be
described as a smooth cylinder with its radius and height
represented by $r$ and $h$. Then its Hamiltonian can be written
as:

\begin{equation}
\label{eq1}
H^S = \frac{1}{2}m\dot {r}^2 + \frac{1}{2}m\omega _{RBM} ^2r^2.
\end{equation}

Here the superscript $S$ means the SWNT. $m = \left( {2\pi rh}
\right)n_\sigma m_c $ is the mass of the cylinder, $n_\sigma $ is
the mean surface density of carbon atoms on the tube wall, and
$m_c $ is the mass of carbon atom. $\dot {r}$ is the time
derivative of the radius, and $\omega _{RBM} $ is the vibration
frequency of the RBM mode. It is well known that the RBM frequency
of the SWNT satisfies a simple relationship with the nanotube
radius,

\begin{equation}
\label{eq2}
\omega _{RBM} = D / \left( {2r_0 } \right),
\end{equation}

\noindent where $D$ keeps a constant for any kind of SWNTs and
$r_0 $ is the equilibrium radius of the SWNT. Thus we have

\begin{equation}
\label{eq3}
H^S = \frac{1}{2}m\dot {r}^2 + \frac{1}{8}mD^2\left( {\frac{r}{r_0 }}
\right)^2.
\end{equation}

For the DWCNTs, the Hamiltonian can be further written as
\cite{r13}:

\begin{equation}
\label{eq4}
H^D = H_{out}^S + H_{in}^S + V_{io} ,
\end{equation}

\noindent where $H_{out}^S $ and $H_{in}^S $ are the Hamiltonians
of the outer and inner SWNTs, respectively, with their radii as
$r_1 $ and $r_2 $. And $V_{io} $ is the interaction potential
between the inner and outer tubes.

In this work, we assume the interaction between inner- and
outer-tube atoms can be described by the Lennard-Jones(LJ)
potential,

\begin{equation}
\label{eq5}
u\left( x \right) = 4\varepsilon \left[ { - \left( {\frac{\sigma }{x}}
\right)^6 + \left( {\frac{\sigma }{x}} \right)^{12}} \right],
\end{equation}

\noindent with parameters, $\varepsilon = 2.964$ meV and $\sigma =
3.407$ {\AA} \cite{r12}. Based on our previous work \cite{r26},
when the tube height is large enough, the interaction potential
can be writtend as

\begin{eqnarray}
\label{eq6} V_{io} = &&N_{in} \ast 3\pi r_1 \varepsilon n_\sigma
\ast V_{KERNEL} \nonumber\\
&&+ N_{out} \ast 3\pi r_2 \varepsilon n_\sigma \ast V_{KERNEL} ,
\end{eqnarray}

\noindent
where

\begin{equation}
\label{eq7} V_{KERNEL} = \left[ { - \frac{\sigma ^6}{\left( {4r_1
r_2 } \right)^{\frac{5}{2}}}I_5 + \frac{21}{32}\frac{\sigma
^{12}}{\left( {4r_1 r_2 } \right)^{\frac{11}{2}}}I_{11} } \right],
\end{equation}

\begin{equation}
\label{eq8} I_{2k+1} = \int_0^{\frac{\pi}{2}} {\frac{dt}{\left(
{a^2 + \sin ^2t} \right)^{\frac{2k+1}{2}}}} , \quad a^2 =
\frac{\left( {r_1 - r_2 } \right)^2}{4r_1 r_2 }.
\end{equation}

$N_{in} = 2\pi r_2 n_\sigma h$ and $N_{out} = 2\pi r_1 n_\sigma h$
are the number of atoms of the inner and outer tube segment,
respectively. $k$ is an integer.

For the DWCNTs, when inner tube radius $r_2 $ is larger than
$(r_{1}-r_{2}) \approx  3.4$ {\AA}, the $a^2 \le \frac{1}{8}$, so
we can use our series expression \cite{r26} for the Eq.
(\ref{eq8}), and get

\begin{widetext}

\begin{eqnarray}
\label{eq9} I_{2k + 1} \approx \left\{ {\frac{1}{\left( {2k - 1}
\right)!!}\left( {\frac{2}{a^2}} \right)^k\sum\limits_{m = 0}^{k -
1} {\frac{\left[ {\left( {2m} \right)!} \right]^2}{\left[ {m!}
\right]^3}\frac{\left( {k - m - 1} \right)!}{2}\left(
{\frac{a}{4}} \right)^{2m}} + \frac{\left( {2k - 1}
\right)!!}{\left( {2k} \right)!!}} \right\}.
\end{eqnarray}

Now we can obtain the dynamical matrix analytically,

\begin{equation}
\label{eq10} F = \left[ {{\begin{array}{*{20}c}
 {\frac{1}{m_{out} }\frac{\partial ^2H^D}{\partial r_1 ^2}} \hfill &
{\frac{1}{\sqrt {m_{out} m_{in} } }\frac{\partial ^2H^D}{\partial r_1
\partial r_2 }} \hfill \\
 {\frac{1}{\sqrt {m_{in} m_{out} } }\frac{\partial ^2H^D}{\partial r_2
\partial r_1 }} \hfill & {\frac{1}{m_{in} }\frac{\partial ^2H^D}{\partial
r_2^2 }} \hfill \\
\end{array} }} \right] = \left[ {{\begin{array}{*{20}c}
 {\left( {\omega _{RBM}^{out} } \right)^2} \hfill & 0 \hfill \\
 0 \hfill & {\left( {\omega _{RBM}^{in} } \right)^2} \hfill \\
\end{array} }} \right] + \left[ {{\begin{array}{*{20}c}
 {K_{11} } \hfill & {K_{12} } \hfill \\
 {K_{21} } \hfill & {K_{22} } \hfill \\
\end{array} }} \right] \quad .
\end{equation}

Here,
\begin{subequations}
\label{eq11}
\begin{equation}
K_{11}  = \frac{1}{m_{out} }\frac{\partial ^2V_{io} }{\partial
r_1^2 } \approx C_0 \left[ {\frac{88}{3}s_0^{12} -
\frac{40}{3}s_0^6 - \left( {\frac{44}{3}s_0^{11} -
\frac{20}{3}s_0^5 } \right)s_1 - \left( {\frac{33}{10}s_0^{10} -
\frac{5}{4}s_0^4 } \right)s_1^2 + \ldots } \right],\label{eq11a}
\end{equation}
\begin{equation}
K_{21} = K_{12} = \frac{1}{\sqrt {m_{out} m_{in} } }\frac{\partial
^2V_{io} }{\partial r_1 \partial r_2 } \approx C_0 \left[ { -
\frac{88}{3}s_0^{12} + \frac{40}{3}s_0^6 + \left( { -
\frac{11}{30}s_0^{10} + \frac{5}{12}s_0^4 } \right)s_1^2 + \ldots
} \right],\label{eq11b}
\end{equation}
\begin{equation}
K_{22} = \frac{1}{m_{in} }\frac{\partial ^2V_{io} }{\partial r_2^2
} \approx C_0 \left[ {\frac{88}{3}s_0^{12} - \frac{40}{3}s_0^6 +
\left( {\frac{44}{3}s_0^{11} - \frac{20}{3}s_0^5 } \right)s_2 -
\left( {\frac{33}{10}s_0^{10} - \frac{5}{4}s_0^4 } \right)s_2^2 +
\ldots } \right],\label{eq11c}
\end{equation}
\end{subequations}

\end{widetext}

 where $s_0=\frac{\sigma}{d_0}$, $s_1=\frac{\sigma}{r_1}$, and $s_2=\frac{\sigma}{r_2}$.
$d_0 = r_1 - r_2 $ is the distance between the inner and outer
tubes. Above formulae are suitable for any tube-tube distance.

Assuming $d_0 = 3.4$ \AA $ \approx \sigma $, which is equal to the
interlayer spacing of graphite, we can further get

\begin{subequations}
\label{eq12}
\begin{eqnarray}
K_{11}&&=\frac{1}{m_{out} }\frac{\partial ^2V_{io} }{\partial
r_1^2} \nonumber\\
&&\approx C_0 \left[ {16 - 8s_1 - \frac{41}{20}s_1^2 -
\frac{41}{40}s_1^3 + \ldots } \right],\label{eq12a}
\end{eqnarray}
\begin{eqnarray}
K_{21}&&= K_{12} = \frac{1}{\sqrt {m_{out} m_{in} }
}\frac{\partial ^2V_{io} }{\partial r_1 \partial r_2 } \nonumber
\\
&&\approx C_0 \left[ { - 16 + \frac{1}{20}s_1^2 +
\frac{1}{20}s_1^3 + \ldots } \right],\label{eq12b}
\end{eqnarray}
\begin{eqnarray}
K_{22} && = \frac{1}{m_{in} }\frac{\partial ^2V_{io} }{\partial
r_2^2 } \nonumber\\
&&\approx C_0 \left[ {16 + 8s_2 - \frac{41}{20}s_2^2 +
\frac{41}{40}s_2^3 + \ldots } \right].\label{eq12c}
\end{eqnarray}
\end{subequations}

Here $C_0 = \frac{3\pi n_\sigma \varepsilon }{m_c }$. Choosing
$n_\sigma = \frac{4}{3\sqrt 3 }\frac{1}{a_{C - C}^2 }$, with $a_{C
- C} $ the bond length of the SWNT, being about 1.42 \AA,
$\varepsilon = 2.964$ meV, and $m_c = 12.011$ amu, we can obtain
$C_0 \approx 241 $ {cm}$^{ - 2}$. In this work, the prefactor $D$
in Eq. {\ref{eq2}} is always chosen as 225 cm$^{ - 1}$nm. By
diagonalizing the dynamical matrix of Eq. 10, the RBLM frequencies
can be obtained.

For example, the RBLM frequencies of (5, 5)@(10, 10) DWCNT are
calculated using above $C_0 $ and $D$. Then one can easily obtain
$\omega _{LF} = 173.9$ cm$^{ - 1}$ and $\omega _{HF} = 340.8$
cm$^{ - 1}$. And the corresponding RBM frequencies of (5, 5) and
(10, 10) SWNTs obtained by Eq. {\ref{eq2}} are $\omega _{\left(
{5,5} \right)} = 331.9$ cm$^{ - 1}$ and $\omega _{\left( {10,10}
\right)} = 165.9$ cm$^{ - 1}$. Thus the frequency shifts of the
RBLMs in the DWCNT are $\Delta \omega _{LF}^{LJ} = \omega _{LF} -
\omega _{\left( {10,10} \right)} = 8.0$ cm$^{ - 1}$ and $\Delta
\omega _{HF}^{LJ} = \omega _{HF} - \omega _{\left( {5,5} \right)}
= 8.9$ cm$^{ - 1}$, respectively.

To make a comparison, a first-principle calculation has been
performed on the (5, 5)@(10, 10) DWNT to check the validity of our
method. A supercell geometry was adopted so that the DWCNT is
aligned in a hexagonal array with nearby DWCNT center distance of
25 {\AA}, which is found to be larger enough to prevent the
tube-tube interactions. The K-points sampling in the reciprocal
space is a uniform grid $1\times 1\times 24$ along the nanotube
axis ($Z$ direction) in our calculations. After structure
relaxation on both of the lattice constant along the tube axis and
the atomic positions, the optimal structure is obtained when the
residue forces acting on all the atoms were less than 0.01
eV/{\AA}. The final dynamical matrix is constructed using cumulant
force-constant (CFC) method \cite{r27, r28}. Because of the
advantage of the CFC method in dealing with the low frequency part
of the phonon dispersion, only a small supercell $\left( {1\times
1\times 2} \right)$ is sufficient to calculate accurately the
phonon modes on the $\Gamma $ point and the $X$ point. Our
\textit{ab initio} calculations were performed using highly
accurate projected augmented wave (PAW) method \cite{r29},
implemented in the Vienna \textit{ab initio} simulation package
(VASP) package \cite{r30}. They are based on the
density-functional theory in the local-density approximation
(LDA).

\begin{figure}[htbp]
\caption{\label{fig1} (Color online) Calculated phonon dispersion
and the density of states for the (5, 5)@(10, 10) are shown in the
left and right panels, respectively.}
\includegraphics[width=3.5in,height=4.36in]{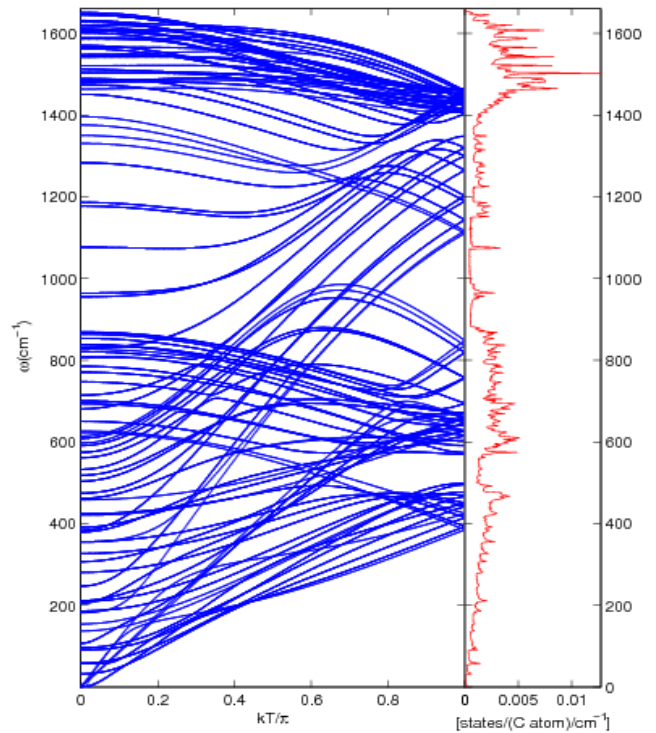}
\end{figure}

The obtained phonon dispersion curves and the density of states
for the (5, 5)@(10, 10) DWCNT are shown in Fig. 1, in which the
low-frequency dispersions are very well reproduced. Its RBLM
frequencies can be determined by analysis of the normal modes,
which are $\omega _{LF}^{ab} = 182.2$ cm$^{ - 1}$ and $\omega
_{HF}^{ab} = 354.6$ cm$^{ - 1}$. The RBM frequencies of (5, 5) and
(10, 10) SWNTs are also calculated by the same \textit{ab initio}
method under the same calculation conditions as the DWCNT, which
are $\omega _{\left( {5,5} \right)}^{ab} = 343.8$ cm$^{ - 1}$ and
$\omega _{\left( {10,10} \right)}^{ab} = 173.6$ cm$^{ - 1}$. Then,
the corresponding frequency shifts of them in the DWCNT are
$\Delta \omega _{LF}^{ab} = \omega _{LF}^{ab} - \omega _{\left(
{10,10} \right)}^{ab} = 8.6$ cm$^{ - 1}$ and $\Delta \omega
_{HF}^{ab} = \omega _{HF}^{ab} - \omega _{\left( {5,5}
\right)}^{ab} = 10.8$ cm$^{ - 1}$, respectively. These results are
consistent with our above obtained results. The residual
difference can be further minimized by adopting the
first-principle tube radii of $R_{\left( {5,5} \right)} = 3.411$
{\AA} and $R_{\left( {10,10} \right)} = 6.766$ {\AA}, and the
first-principle RBM frequencies of the SWNTs in the dynamical
matrix of Eq. 10. By this way, the new frequency shifts obtained
by our analytical method will be $\Delta \omega _{LF}^{LJ} = 9.0$
cm$^{ - 1}$ and $\Delta \omega _{HF}^{LJ} = 10.3$ cm$^{ - 1}$,
agreeing very well with above first-principle values.

Our approach can be further applied to the multi-walled carbon
nanotubes (MWCNTs). Here we take a 3-layer MWCNT as an example.
Again, we can write the system Hamiltonian as

\begin{equation}
\label{eq13}
H^T = \sum\limits_{i = 1}^3 {H_i^S } + \sum\limits_{i \ne j} {V_{ij} } ,
\end{equation}

\noindent where $H_i^S $ is the Hamiltonian of the $i$th layer,
and $V_{ij} $ is the interaction potential between the $i$th and
the $j$th layer nanotubes. And the radius of the $i$th layer is
represented by $r_i $ with $r_1 > r_2 > r_3 $. So, its dynamical
matrix can be written as:

\begin{eqnarray}
\label{eq14}
 F = &&\left[ {{\begin{array}{*{20}c}
 {\frac{1}{m_1 }\frac{\partial ^2H^T}{\partial r_1^2 }} \hfill &
{\frac{1}{\sqrt {m_1 m_2 } }\frac{\partial ^2H^T}{\partial r_1 \partial r_2
}} \hfill & {\frac{1}{\sqrt {m_1 m_3 } }\frac{\partial ^2H^T}{\partial r_1
\partial r_3 }} \hfill \\
 {\frac{1}{\sqrt {m_2 m_1 } }\frac{\partial ^2H^T}{\partial r_2 \partial r_1
}} \hfill & {\frac{1}{m_2 }\frac{\partial ^2H^T}{\partial r_2^2 }} \hfill &
{\frac{1}{\sqrt {m_2 m_3 } }\frac{\partial ^2H^T}{\partial r_2 \partial r_3
}} \hfill \\
 {\frac{1}{\sqrt {m_3 m_1 } }\frac{\partial ^2H^T}{\partial r_3 \partial r_1
}} \hfill & {\frac{1}{\sqrt {m_3 m_2 } }\frac{\partial ^2H^T}{\partial r_3
\partial r_2 }} \hfill & {\frac{1}{m_3 }\frac{\partial ^2H^T}{\partial r_3^2
}} \hfill \\
\end{array} }} \right] \nonumber\\
 = &&\left[ {{\begin{array}{*{20}c}
 {\left( {\omega _{RBM}^1 } \right)^2} \hfill & 0 \hfill & 0 \hfill \\
 0 \hfill & {\left( {\omega _{RBM}^2 } \right)^2} \hfill & 0 \hfill \\
 0 \hfill & 0 \hfill & {\left( {\omega _{RBM}^3 } \right)^2} \hfill \\
\end{array} }} \right] \nonumber\\
&&+ \left[ {{\begin{array}{*{20}c}
 {K_{11}^{12} + K_{11}^{13} } \hfill & {K_{12}^{12} } \hfill & {K_{13}^{13}
} \hfill \\
 {K_{21}^{12} } \hfill & {K_{22}^{12} + K_{22}^{23} } \hfill & {K_{23}^{23}
} \hfill \\
 {K_{31}^{13} } \hfill & {K_{32}^{23} } \hfill & {K_{33}^{13} + K_{33}^{23}
} \hfill \\
\end{array} }} \right],
\end{eqnarray}

\noindent where $K_{\alpha \beta }^{ij} $ means the $\left(
{\alpha ,\beta } \right)$ element of the $K$ matrix, which comes
from the interaction between the $i$th and the $j$th layer tubes.
If condition $a_{13}^2 = \frac{\left( {r_1 - r_3 } \right)^2}{4r_1
r_3 } \ll 1$ can be satisfied, the Eq. (\ref{eq11}) can also be
used to get the force constants between the 1st and the 3rd layer
tubes. So using Eq. (\ref{eq11}), the force constants between any
two different layer tubes can be obtained easily. Here, for the 3
layers MWNT, we use $d_0 = 6.8$ {\AA } as the distance between the
1st and the 3rd layer tubes, then Eq. (\ref{eq14}) is fully
determined.

To examine the precision of our result, we compare them with other
numerical ones. The RBLM frequencies as a function of the outer
tube radius for $\left( {n,n} \right)@\left( {n + 5,n + 5}
\right)@\left( {n + 10,n + 10} \right)$ MWCNTs are plotted in Fig.
2 (For comparison, please see figure 5 of Ref. \onlinecite{r12}).
From Fig. 2, it is found that our results match Ref.
\onlinecite{r12} very well, and the residual difference may come
from the choice of some parameters, e.g., $D$, $d_0 $ and
$n_\sigma $. In fact, if we only slightly adjust both of $D$ and
$C_0 $, our RBLM frequencies can match all the other publications
\cite{r5,r6,r7,r8,r9,r10,r12} perfectly.

\begin{figure}[htbp]
\caption{\label{fig2} (Color online) Dependence of the RBLM
freqeuncies on the outermost tube radius for 3 layer MWCNTs, i.e.,
$\left( {n,n} \right)@\left( {n + 5,n + 5} \right)@\left( {n +
10,n + 10} \right)$. The highest (HF), the middle (MF) and the
lowest (LF) RBLM are plotted.}
\includegraphics[width=3.5in,height=2.69in]{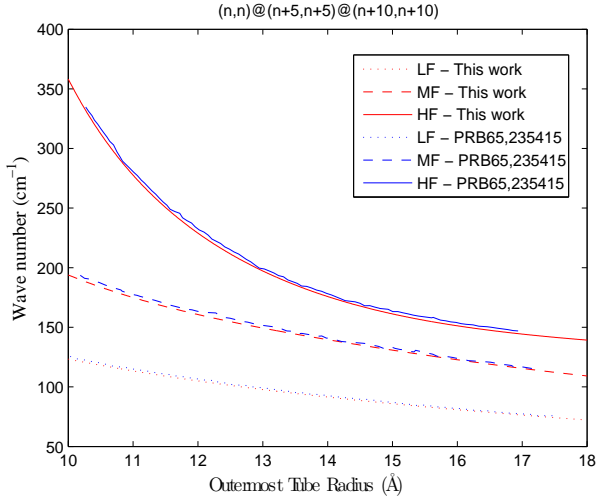}
\end{figure}

Furthermore, we show the dependence of the RBLM freqeuncies on the
number of layers in Fig. 3, in which the radius of the innermost
tube is fixed at 30{\AA}. This result matches the Fig. 6 of Ref.
\onlinecite{r12} perfectly. Based upon Fig. 3, a new method to
determine the number of layers for the MWCNTs can be proposed,
i.e., if the position of the lowest RBLM for a MWNT is obtained,
one can easily find out how many layers are present in the MWCNT
by comparing the experimental data with Fig. 3.

\begin{figure}[htbp]
\caption{\label{fig3}(Color online) Dependence of the RBLM
freqeuncies on the layer number. Here, the radius of the innermost
tube is fixed at 30 \AA.}
\includegraphics[width=3.5in,height=4.72in]{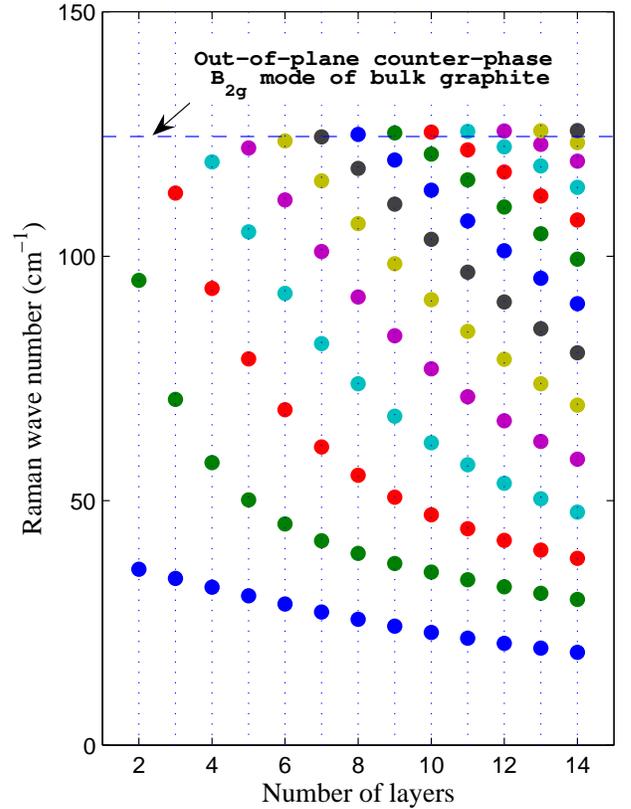}
\end{figure}

Now we would like to illustrate the application of our approach.
In Ref. \onlinecite{r11}, the Raman investigation of the DWCNTs
prepared from thermal conversion of C$_{60}$ encapsulated in the
SWCNTs has been reported, in which the Eq. (\ref{eq2}), suitable
only to the SWCNTs, was used to obtain the tube radii. Now we used
our approach to determine the inner tube radius and the tube-tube
distance. Here we choose a pair of experimental data from Ref.
\onlinecite{r11}, $\omega _{LF} = 150$ cm$^{ - 1}$, and $\omega
_{HF} = 267$ cm$^{ - 1}$. By Eq. (\ref{eq11}), the high and low
RBLM frequencies for different $r_2 $ can be obtained at the
certain $d_0 $. Then comparing them with the experimental
frequencies, we can get two different inner tube radii. In a real
system, these two inner tube radii value must be the same, and so
we have to adjust the distance $d_0 $ between the inner and outer
tubes. Finally, a most reliable inner tube radius can be found. In
Fig. 4, we finally set $d_0 $ to be 3.558 \AA, making the cross
points of low and high frequencies with the experimental
frequencies lie on one vertical line, from which a reliable inner
tube radius of about 4.453 {\AA} is obtained, i.e., $R_{in} =
4.453$ \AA, and so $R_{out} = 8.011$ \AA.

\begin{figure}[htbp]
\caption{\label{fig4}(Color online) The Raman frequency vs the
inner tube radius, which is obtained after adjusting the distance
$d_0 $ between the inner and outer walls to about 3.558 {\AA}. }
\includegraphics[width=2.96in,height=4.72in]{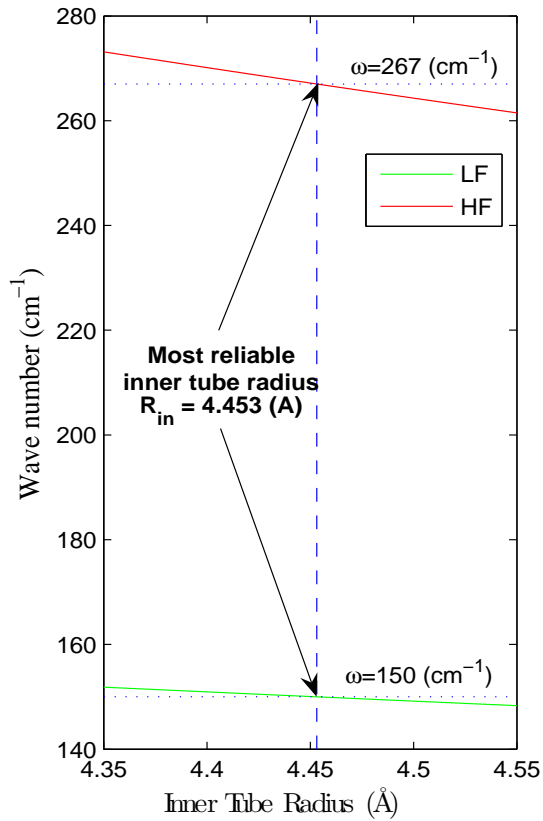}
\end{figure}

In summary, we presented an analytically solvable approach for the
radial breathing-like phonon modes of the MWNTs, where the tube
walls are treated as coupled oscillators, and the carbon atoms on
nanotubes are distributed continuously. Then the force constants
between inner and outer tubes, and the relationship of the RBLM
frequencies with the tube radii can be obtained analytically. It
is found that our obtained RBLM frequencies are well consistent
with those in other publications and experiments. Finally, our
result can be helpful to determine the radii of the DWCNTs and
MWCNTs.

\begin{acknowledgments}
The authors acknowledge support from the Natural Science
Foundation of China under Grants No. 10474035 and No. A040108. The
authors also thanks support to this work from a Grant for State
Key Program of China through No. 2004CB619004. Our first-principle
calculations have been done on the Sun Fire V20z computers.

\end{acknowledgments}

\end{document}